\documentclass[pdftex,12pt]{article}

\usepackage{amsmath,amssymb,amsfonts} 
\usepackage[pdftex]{graphicx}                 
\usepackage{color}                    
\usepackage{hyperref}                 
\usepackage[english]{babel}
\usepackage{fancyhdr} 
\usepackage{longtable} 
\usepackage{astron}
\usepackage{pdfpages}
\usepackage{listings}

\definecolor{dkgreen}{rgb}{0,0.6,0}
\definecolor{gray}{rgb}{0.5,0.5,0.5}
\definecolor{mauve}{rgb}{0.58,0,0.82}
\usepackage[vcentering,dvips]{geometry}
\begin{document} 

\begin{Huge}
\begin{center}
\vspace{3cm}
\textbf{JPARSEC:\\}
\vspace{0.5cm}
\textbf{a Java package for astronomy with twelve years of development and use}
\\ 
\vspace{1cm}
\end{center}
\end{Huge}
\thispagestyle{empty}
\begin{Large}
\begin{center}\textbf{Abstract}\end{center}
   JPARSEC is a Java library initially developed to implement a complete set of algorithms to compute ephemerides. The project started twelve years ago and soon evolved to also cover astrophysical modeling and outreach. JPARSEC itself includes more than 250 000 lines of code, similar or larger than other popular tools like Astropy, complemented with models and other projects operational since years. The library is focused on being robust, optimized, well written and documented, providing a complete set of tools for astronomers, with special 
   attention to radioastronomy. The projects written using JPARSEC as the core library cover a wide range of fields from models with complete graphical user interfaces to an ephemerides server with high quality charts, or even an Android planetarium. In this Paper I present the main characteristics of JPARSEC, with special attention to the core library 
   and documentation, and I describe the main projects developed based on it.

\end{Large}

\vspace{1cm}

\begin{center}
\begin{Large}
\textbf{
Tom\'as Alonso Albi}
\end{Large}
\\ 
\vspace{0.5cm}
Observatorio Astron\'omico Nacional (IGN)\\
Calle Alfonso XII, 3, 28014 Madrid, Spain \\
t.alonso@oan.es\\

\end{center}

\bibliographystyle{astron}
%
%
%


\def\jnl@style{\it}
\def\aaref@jnl#1{{\jnl@style#1}}

\def\aaref@jnl#1{{\jnl@style#1}}

\def\aj{\aaref@jnl{AJ}}                   
\def\araa{\aaref@jnl{ARA\&A}}             
\def\apj{\aaref@jnl{ApJ}}                 
\def\apjl{\aaref@jnl{ApJ}}                
\def\apjs{\aaref@jnl{ApJS}}               
\def\ao{\aaref@jnl{Appl.~Opt.}}           
\def\apss{\aaref@jnl{Ap\&SS}}             
\def\aap{\aaref@jnl{A\&A}}                
\def\aapr{\aaref@jnl{A\&A~Rev.}}          
\def\aaps{\aaref@jnl{A\&AS}}              
\def\azh{\aaref@jnl{AZh}}                 
\def\baas{\aaref@jnl{BAAS}}               
\def\jrasc{\aaref@jnl{JRASC}}             
\def\memras{\aaref@jnl{MmRAS}}            
\def\mnras{\aaref@jnl{MNRAS}}             
\def\pra{\aaref@jnl{Phys.~Rev.~A}}        
\def\prb{\aaref@jnl{Phys.~Rev.~B}}        
\def\prc{\aaref@jnl{Phys.~Rev.~C}}        
\def\prd{\aaref@jnl{Phys.~Rev.~D}}        
\def\pre{\aaref@jnl{Phys.~Rev.~E}}        
\def\prl{\aaref@jnl{Phys.~Rev.~Lett.}}    
\def\pasp{\aaref@jnl{PASP}}               
\def\pasj{\aaref@jnl{PASJ}}               
\def\qjras{\aaref@jnl{QJRAS}}             
\def\skytel{\aaref@jnl{S\&T}}             
\def\solphys{\aaref@jnl{Sol.~Phys.}}      
\def\sovast{\aaref@jnl{Soviet~Ast.}}      
\def\ssr{\aaref@jnl{Space~Sci.~Rev.}}     
\def\zap{\aaref@jnl{ZAp}}                 
\def\nat{\aaref@jnl{Nature}}              
\def\iaucirc{\aaref@jnl{IAU~Circ.}}       
\def\aplett{\aaref@jnl{Astrophys.~Lett.}} 
\def\apspr{\aaref@jnl{Astrophys.~Space~Phys.~Res.}}
\def\bain{\aaref@jnl{Bull.~Astron.~Inst.~Netherlands}} 
\def\fcp{\aaref@jnl{Fund.~Cosmic~Phys.}}  
\def\gca{\aaref@jnl{Geochim.~Cosmochim.~Acta}}   
\def\grl{\aaref@jnl{Geophys.~Res.~Lett.}} 
\def\jcp{\aaref@jnl{J.~Chem.~Phys.}}      
\def\jgr{\aaref@jnl{J.~Geophys.~Res.}}    
\def\jqsrt{\aaref@jnl{J.~Quant.~Spec.~Radiat.~Transf.}}
\def\memsai{\aaref@jnl{Mem.~Soc.~Astron.~Italiana}}
\def\nphysa{\aaref@jnl{Nucl.~Phys.~A}}   
\def\physrep{\aaref@jnl{Phys.~Rep.}}   
\def\physscr{\aaref@jnl{Phys.~Scr}}   
\def\planss{\aaref@jnl{Planet.~Space~Sci.}}   
\def\procspie{\aaref@jnl{Proc.~SPIE}}   

\let\astap=\aap
\let\apjlett=\apjl
\let\apjsupp=\apjs
\let\applopt=\ao

\newcommand\Msun{M$_{{\odot}}$\:}
\newcommand\pow[1]{10$^{#1}$}
\newcommand\Rin{R$_{\mathrm{in}}$\:}
\newcommand\Rout{R$_{\mathrm{out}}$\:}
\newcommand\kms{km~s$^{-1}$}
\newcommand\Lsun{L$_{{\odot}}$}






\newpage
\tableofcontents

\pagestyle{fancy}



\normalsize

\section{Introduction}

One of the tasks of most astronomers is to develop a comfortable set of tools to improve the efficiency and quality of the scientific research. Choosing the right means, for instance an adequate programming language and development strategy, is crucial to maximize long-term productivity. But these choices are subjective, usually dependent on the knowledge, the goals, and the effort the astronomer is willing to carry out, and also time dependent, since tools evolve with time and eventually new better ones can appear. Developing large programs can be exciting, but time consuming, leading to a decrease in scientific productivity unless most of the code can be reused.

Considering these points I started in 2006 a project called JPARSEC, using the Java programming language. By then this language was already at the top in the TIOBE language popularity index\footnote{https://www.tiobe.com/tiobe-index/}, and has been at the top since then most of the years, sometimes exceeded by C, but still way ahead of most other languages. There are reasons for this and I will enumerate the main ones.

Java is extremely productive. Since it is an object oriented language, it offers a high degree level of abstraction to implement simple or complex programs and to keep them easy to maintain and scale. There are excellent free tools to develop in this language, the so called Integrated Development Environments (IDEs), that offer among many other useful features syntax/error corrections or direct access to the extensive documentation of the Java language. Both the object oriented programming and the use of IDEs requires some initial effort, but in the long term the productivity and safety of the work (the capability to do more in less time with a minimal error rate) is immensely increased.

Java is a cross-platform and cross-architecture language, which means that a Java program, correctly coded, will run without the need of recompilation or system dependent libraries across different machines and operating systems, giving the same mathematical results on them up to the last digit. It is also incremental, which means that a program written and compiled for a given version will run in later versions of Java. These features are of great importance for scientific research, and prevent issues with the models or charts when evolving toward newer operating systems. The same applies to most external libraries freely available, some of them directly integrated in JPARSEC.

Java programs are very fast, specially compared to dynamically typed and interpreted languages like Python, than can be many times slower compared to Java, Fortran, or C in mathematical operations\footnote{https://modelingguru.nasa.gov/docs/DOC-2676 includes some speed tests for different languages, showing that in some cases Python is 100 times slower than other languages. Without using external specific libraries, Java lags behind only in matrix operations.}. To implement in an interpreted language some of the applications of JPARSEC presented in this Paper would be impossible or hardly feasible, since one of the critical points in any algorithm is testing, and a thorough testing requires many executions. A slow language limits the complexity a given algorithm can reach, at least with a reasonable safety level. Java also offers an excellent profiler called VisualVM\footnote{https://visualvm.github.io/}, extremely useful to improve the performance of Java programs.

There are other comparable tools available, mainly Astropy (\cite*{astropy2013}). JPARSEC has been developed with a different strategy (not community oriented, at least until reaching a mature and stable level), so I will not do any technical comparison. There are clear differences: JPARSEC is not focused only in science, but tries to cover a wider range of applications, for instance it can be useful for outreach. But since the entire project has been developed by only one astronomer there is probably some lack of specific features for different fields in astronomy.

In this Paper I present the JPARSEC library, first publicly released ten years ago, but developed and optimized over twelve years using an homogeneous and strict code style and strategy. First I start with a general picture of the library and other related tools, then I describe with some detail the main features of JPARSEC, and later in Section 4 I present some elaborated projects based on it, that in some cases have required thousands of hours or work. Documentation is also extensive, including a complete web tutorial describing how to download and start using JPARSEC that is presented before concluding.


\section{Organization and packages}

\begin{figure}
    \centering
    \includegraphics[width=0.75\textwidth]{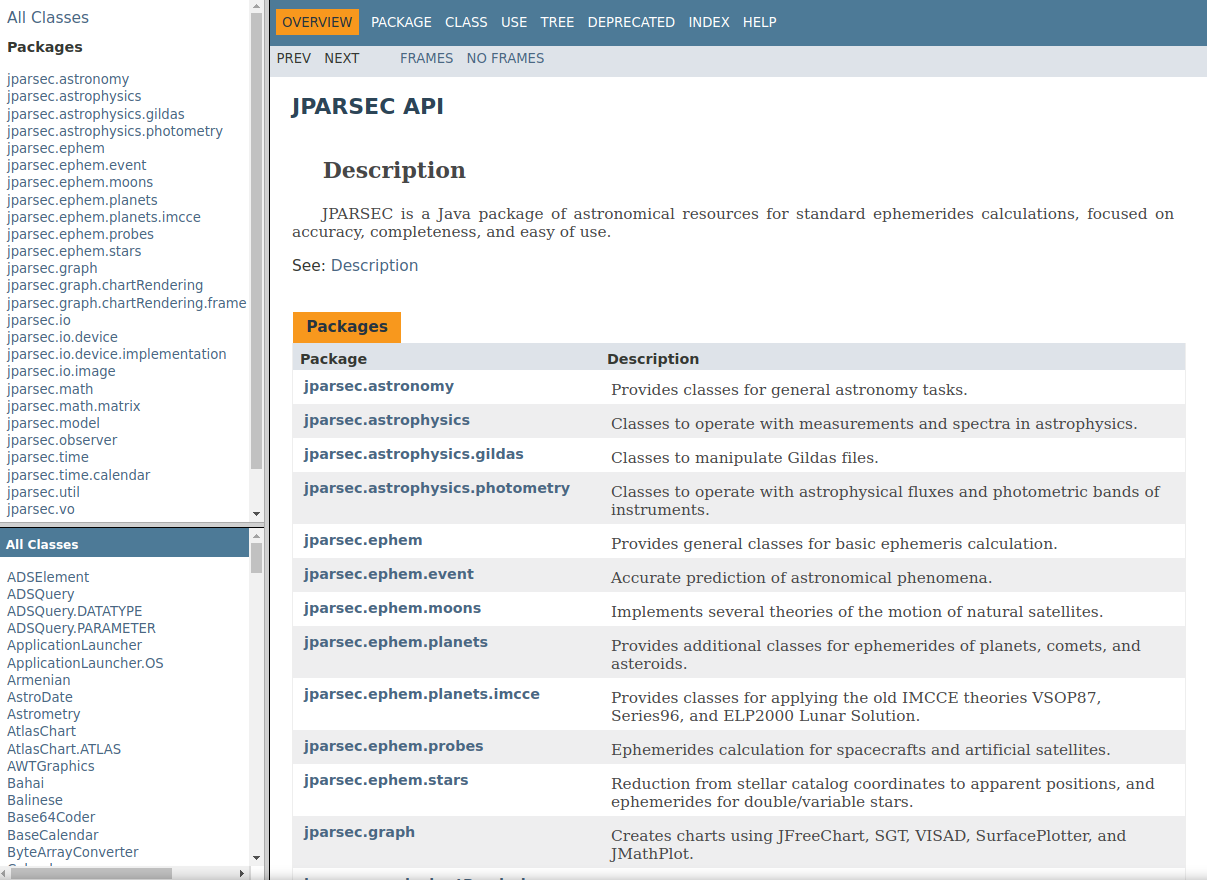}
    \caption{The JPARSEC API documentation is available as an interactive and extensive web page. On the top-left corner the image shows the list of packages.}
    \label{fig:doc}
\end{figure}

JPARSEC is structured in 26 directories (11 main packages and 15 additional sub-packages), containing 274 source files (called Java classes) with 260 000 lines of code released under the GPL license. Documentation on the methods is included inside the code, including information related to how to properly call each function, the units expected for the input and output values, or references to more than a hundred Papers in which ephemerides and other algorithms are based. More than 40 000 lines of code belong to this documentation, that can be accessed through web pages\footnote{http://conga.oan.es/\%7Ealonso/jparsec/doc/} thanks to the Javadoc tool (see Fig. \ref{fig:doc}), or directly with the IDE while writing the code as the documentation of the Java language itself. In addition, there are another 6000 lines of comments dedicated to clarify some decisions taken in the implementation of the algorithms. They are only accessible inside the source code.

JPARSEC has evolved during the years to agglutinate features and interrelate all them. The priority has been always code quality and documentation, following the standard conventions for the Java language\footnote{http://www.oracle.com/technetwork/java/codeconvtoc-136057.html}. When the core library was solid enough, other projects based on it were developed. In practice, JPARSEC passed to production phase with the development of other tools, and soon it was evident that some very specific features integrated inside the core library should be moved to external tools to maintain JPARSEC as a clean common core. For practical purposes and faster development, the strict code conventions and organization was maintained only for the library, not for the external tools.

\section{Features of JPARSEC}

In this section I present the main practical features of the JPARSEC library, avoiding unnecessary details. Many of them were developed around ten years ago, and progressively JPARSEC has become more stable, evolving mostly for bug fixes in the last years.

\subsection{Planetary ephemerides}

Computing ephemerides in JPARSEC requires three objects: one representing the observer (with properties such as geographical longitude, latitude, elevation, or even the ambient temperature and pressure of the site that could affect the refraction), another representing the time, with different fields for the date and time and the time scale, and another with the properties of the ephemerides to be computed. These objects are also used for many other purposes in JPARSEC, and their implementations are located in the packages jparsec.observer, jparsec.time, and jparsec.ephem. To help to construct the observer JPARSEC contains a database with more than 4000 cities around the world and 1200 observatories, although the observer can also be located anywhere in the solar system, for instance at a given position in the space or on the surface of Mars or any other body. A database of planetary features is integrated in JPARSEC for this purpose\footnote{http://planetarynames.wr.usgs.gov/}. To construct the date it is possible to use the Julian day number as a double value, but also as a BigDecimal object to store it with a precision in the nanosecond level. The supported time scales are local time, Universal Time UT1 and UTC, Terrestrial Time, and Barycentric Dynamical Time. Dates can also be obtained from many calendars\footnote{JPARSEC implements the algorithms described in the book Calendrical Calculations, by N. Dershowitz and E. M. Reingold.}, and it is easy to transform dates between different calendars or time scales.

The implementation of the ephemerides object and the rest of algorithms to compute the positions are located in the jparsec.ephem package and their sub-packages. The most relevant properties of the ephemerides object are the selection of the specific algorithm to compute the coordinates for the object, and the set of algorithms to be used for reducing these coordinates to calculate the position of the object as seen by the observer. JPARSEC supports most of the theories published in the literature: the JPL ephemerides (DE200/403/405/406/413/414/422/424/430) (\cite*{Folkner2014}), VSOP87 (\cite*{Bretagnon1988}), ELP2000 (\cite*{Chapront1988}), Series96 (\cite*{Chapront1995}), and the analytical theory by S. L. Moshier\footnote{Program aa-56 available at http://www.moshier.net/}. For natural satellites the implemented algorithms are the theory by \cite{Lainey2007} for the Martian satellites, E2x3, E5 (\cite*{Lieske1987,Lieske1998}) and L1 (\cite*{Lainey2006}) for Jovian satellites, TASS1.7 (\cite*{Vienne1995,Duriez1997}) and the theory by \cite{Dourneau1993} for Saturnian ones, and GUST86 (\cite*{Laskar1987}) for the Uranian satellites. Low precision ephemerides based on orbital elements are also supported for dwarf satellites. All data files required for these computations are integrated in JPARSEC as dependencies, including the JPL ephemerides files for the year range 1900-2100, although the analytical theory by Moshier (a fit to DE404 ephemerides between years 3000 B.D. and 3000 A.D.) can be used instead.

Respect reduction methods, JPARSEC supports the IAU 1980, 2000, 2006, and 2009 resolutions (that established different methods for precession, nutation, obliquity, and Greenwich mean sidereal time), which means that it can reproduce to an accuracy well below the milliarcsecond the ephemerides results of the Astronomical Almanac or the JPL Horizons and IMCCE ephemerides servers\footnote{The implementation of ephemerides in these servers is different and use some old algorithms. Horizons is located at https://ssd.jpl.nasa.gov/horizons.cgi, and the IMCCE server at https://www.imcce.fr/services/ephemerides/}. Ephemerides can be generated for different kind of coordinates (geometric, astrometric, apparent), and referred to any output equinox and frame (ICRF, mean dynamical, FK5, FK4). The code listing \ref{list1} shows a basic example that creates the required objects to compute the position of Saturn.

JPARSEC produce ephemerides with many additional data, including the physical ephemerides, disk orientations (longitude of central meridian, position angle of axis and pole for the different IAU resolutions), or rise and set times, among others. Ephemerides are supported for planets, natural satellites, artificial satellites (implementing the SGP4/SDP4 and SGP8/SDP8 propagation models) and spacecrafts, minor objects, and stars (including orbits of visual binary stars). Among other catalogs the Bright Star Catalog (\cite*{Hoffleit1991}) and Sky2000 Master Catalog (\cite*{Myers2001}) are included as dependencies. JPARSEC also computes the instant of many different kinds of astronomical events, not only the main ones such as eclipses or moon phases, but even more complex ones like the mutual events (mutual eclipses and occultations) between the satellites of Jupiter, Saturn, and Uranus, with a remarkable accuracy when comparing the same computations with other sources. 

The nominal accuracy of the reduction of coordinates performed by JPARSEC is in the 0.1 mas level, and many of the algorithms used are based on SOFA (\cite*{Hohenkerk2012}) and NOVAS (\cite*{Kaplan2012}). But the integration of different theories and the flexibility and completeness of the implementation simplifies the computation of ephemerides that otherwise would be difficult to perform. For instance, one of the code examples distributed with JPARSEC computes solar eclipses produced by Phobos as observed by the Curiosity rover on Mars, showing excellent consistency with the images taken by the rover. Many interesting details and features available are omitted here since the underlying work by other astronomers would require dozens of additional references to Papers and web pages, that can be found in the source code of JPARSEC and on its web page.

\bigskip
\bigskip
\begin{lstlisting}[caption={Code example that shows how to define the objects time, observer, and ephemerides properties to get the position of Saturn, and to convert later this position to galactic coordinates.},captionpos=b,label={list1}]
// Define time, observer, and ephemerides properties
AstroDate astro = new AstroDate(2006, AstroDate.JANUARY, 1);
TimeElement time = new TimeElement(astro, SCALE.LOCAL_TIME);
CityElement city = City.findCity("Madrid");
ObserverElement obs = ObserverElement.parseCity(city);
EphemerisElement eph = new EphemerisElement(TARGET.SATURN, COORDINATES_TYPE.APPARENT, EphemerisElement.EQUINOX_OF_DATE, EphemerisElement.TOPOCENTRIC, REDUCTION_METHOD.IAU_2006, FRAME.ICRF, ALGORITHM.JPL_DE430);
	 
// Get ephemerides and report results
EphemElement saturn = Ephem.getEphemeris(time, obs, eph, true); // Compute also rise/set/transit
ConsoleReport.fullEphemReportToConsole(saturn);

// Transform position to galactic coordinates
LocationElement gal = CoordinateSystem.equatorialToGalactic(saturn.getEquatorialLocation(), time, obs, eph);
System.out.println("Galactic coordinates: "+gal.toString());	
\end{lstlisting}

\subsection{Constants, measures, error propagation and unit conversions}

The package jparsec.math contains Java classes aimed to help with mathematical operations. There are two classes for physical constants, called Constant and CGSConstant. The second one just provides a few constants in CGS units, while the first one provides around 80 useful values, most of them physical (based on CODATA 2014\footnote{https://physics.nist.gov/cuu/Constants/}) and astronomical constants, but also other practical values. There are values like Constant.SECONDS\_PER\_DAY for simple unit conversions, which may seem unnecessary but shows the development strategy behind JPARSEC. The mentioned constant is used many times for converting seconds to days or the opposite. In many libraries the developer would use the numerical value everywhere, but that practice has an important drawback: the interpretation of the code can be confusing when a value cannot be recognized. In addition, it is not strictly correct from the physical point of view, since in the previous example the intention is to convert days to seconds, not to multiply them by a value that, eventually, could be modified or wrongly typed, producing a bug very difficult to find. This is one of the code quality practices used along the entire JPARSEC library, complemented with others like careful selection of variable names, or the use of enum constants instead of numerical values to offer a set of acceptable values for a given parameter. For instance, the ephemerides object contains some enums with a set of valid values to select the target source, planetary ephemerides, or reduction algorithm, that prevent to provide a wrong input value. The target body is selected with the enum TARGET, which has dozens of possible values for planets or natural satellites, for instance TARGET.MARS for the Mars planet, with additional methods to retrieve its equatorial or polar radius, among other data. JPARSEC always uses enums when there is a limited set of valid values for a given parameter, hence in practice these identifiers behave similarly to (symbolic) constants. 

Measures and error propagation are provided by the class MeasureElement, located in the package jparsec.astrophysics. This class lists many physical constants with their uncertainties, that can be used instead of the values in class Constant to consider these uncertainties in error propagation. Instances of MeasureElement are objects that can be multiplied or added to others to propagate their errors using the method of first order partial derivatives (see \cite{ku1966} for a classic review and other methods). The object has obviously a value and an error, but also an optional unit that can be used to convert the measure to other units. JPARSEC supports the conversion between different flux units (magnitudes in different photometric bands, Jy, among others) and also general conversions between compatible units using the package cds.astro available from the CDS\footnote{http://cds.u-strasbg.fr/resources/doku.php?id=downloads}. An example of the later would be to convert parsecs to au. Users can also define their own units with the combination of other basic ones.

A MeasureElement object can be formatted as a string taking into account the significant digits required to express the value and the error with any of the toString() methods. When writing thousands of values in tables for different formats (html or Latex) this automatic formatting routine is extremely useful.

\subsection{Tables, matrices, fittings, and other mathematical operations}

Tables are supported by the class jparsec.astrophysics.Table. The implementation allows to handle 1d, 2d, and 3d tables filled with double values or MeasureElement objects. There are basic operations from adding or multiplying columns to convolution or interpolation around a given point.

The sub-package jparsec.math.matrix contains classes to decompose matrices using different methods and to solve linear systems. The classes are adapted from the JAMA library\footnote{Although there are much better options like ojAlgo, the JAMA library at https://math.nist.gov/javanumerics/jama/ was easy to integrate and provides enough features and decent performance.}, a relatively little library for linear algebra directly integrated inside JPARSEC with some changes to better interrelate those classes with the jparsec.math package. Matrices are extensively used for the implementation of the IAU 2006 resolutions or the Principal Component Analysis method.

There are many other useful classes in the jparsec.math package. The class Derivation implements the Lagrange method and another method based on spline interpolation to compute derivatives, the Integration class implements the midpoint rule, and the Interpolation class implements linear, spline, and Lagrange methods for interpolating. There are three classes for fittings: LinearFit, specialized in linear fitting with a rigorous treatment of errors; GenericFit, useful to fit series of data to any function of three variables given as a string expression; and Regression, which uses the Java scientific library by Michael Thomas Flanagan\footnote{Available from https://www.ee.ucl.ac.uk/\%7Emflanaga/java/} to fit series of data to more complex functions such as Gaussians or polynomials. Other classes are designed to work with vectors, polynomials, complex values, calculate different kind of statistical averages, among others.

The jparsec.math package includes the class FastMath to provide alternative and approximate implementations of different mathematical functions. This class can be extremely useful to rotate coordinates when showing graphical elements on the screen, where a precision of a fraction of pixel is enough. Some of these implementations are hundreds of times faster than the native Java implementation. FastMath is extensively used for approximate rotation of coordinates in different systems (equatorial, ecliptic, horizontal, and galactic) and for sky and planetary rendering.

\subsection{Interoperatibility with GILDAS: spectra, cubes, and pyGildas extension}

The package GILDAS (\cite*{Gildas2013}) is widely used by the community of radioastronomers to reduce and analyze spectra and cubes. The package jparsec.astrophysics.gildas implements reading and writing data in the internal format used in GILDAS for spectra (the 30m format) and cubes (lmv format), as well as many operations equivalent to those provided by GILDAS. The 30m format implemented in JPARSEC is rectricted to the old format, but can read files with thousands of spectra and thousands of channels per spectrum. Both spectra and cubes can be read and written also in fits format.

The class jparsec.astrophysics.gildas.pyGildas can be used to launch scripts for GILDAS and to retrieve the value of some GILDAS variables as output. For this purpose Python is used in background to launch GILDAS, without the need for any other requirement. The methods in this class hide this process, reduced to simply calling the functions with the script to launch given as a string array.

The interactivity with GILDAS and the support for its native formats have important advantages that will be shown below. The scripts that can be launched with GILDAS are very flexible, but not enough to reduce and analyze large amounts of data automatically.

\subsection{FITS and WCS support}

The package jparsec.io.image includes the classes FitsIO, FitsBinaryTable, and WCS. FitsIO uses the external library nom.tam.fits to read files in fits format and manipulate the HDUs, supporting the BZERO and BSCALE flags to scale the data and the different data types allowed in fits files. The FitsBinaryTable class can be used to handle HDUs containing binary and ascii tables.

The WCS class in JPARSEC provides support to the World Coordinate System (\cite*{Greisen2002}). There are two implementations supported: the one provided by JSky (\cite*{Dolensky1999}), and the one based on SkyView (\cite*{McGlynn2015}). The main difference is that the SkyView implementation supports the image distortion parameters, but only for some of the projections available.

\subsection{The Virtual Observatory package}

The package jparsec.vo is designed to interact with web servers. JPARSEC supports queries to Simbad\footnote{http://simbad.u-strasbg.fr/simbad/} (source solver), Vizier\footnote{http://vizier.u-strasbg.fr/viz-bin/VizieR} (queries to catalogs), ADS\footnote{http://adsabs.harvard.edu/} (to retrieve abstracts, bibtex, and complete articles in PDF format), CDS coordinate converter\footnote{http://cds.u-strasbg.fr/resources/doku.php?id=astrocoordinates} (using the cds.astro external library), and images from the Sloan Digital Sky Survey\footnote{http://www.sdss.org/} (SDSS) and SkyView\footnote{https://skyview.gsfc.nasa.gov/current/cgi/titlepage.pl}. Virtual Observatory tables from Vizier can also be manipulated or created. The modeling tools take advantage of these features, in particular SEDFit to automatically download the photometry for a given source.

The jparsec.vo package includes the class SExtractor, that simplifies the use of this source extractor utility (\cite*{Bertin2010}) and to retrieve its output. There are several code examples showing possible applications, and in particular the web tutorial includes a code piece that reads fits images of a globular cluster taken with the Very Large Telescope (VLT) in two bands and returns the Hertzsprung-Russell (HR) diagram of the cluster.

\subsection{Reading, writing, and manipulating data}

The package jparsec.io contains useful classes to manipulate files and interact with the system. The class FileFormatElement is used to describe a text file formatted in fields over different ranges of columns, and the class ReadFormat uses a FileFormatElement object to read a given field name as a string or as a double value. Most catalogs in JPARSEC (stars, deep sky objects, orbital elements in different formats) can be read with the objects of this kind already defined. But the most useful classes are ReadFile and WriteFile, with many methods to read text files (entire files or parts) and images from external files or located within the dependencies. WriteFile can also use a FileFormatElement object to reproduce the format of a text file.

The class FileIO in this package and jparsec.graph.DataSet are also very useful when reading files. There are methods in these classes to sort values and to convert data into different types, that complement FileFormatElement to read and write data in less strict formats. The class ConsoleReport can then be used to write data to the console instead of to a text file. This class also contains methods for a quick report of the fields in some objects used in JPARSEC for ephemerides, and supports formatting text with the syntaxis used in C and Fortran.

The jparsec.io package includes the classes HTMLReport and LATEXReport. These classes can be very useful to create automatic reports in html and latex formats. Most of the methods in both classes have the same name and input parameters (to create paragraphs, tables, or show figures), which means that with a quick change in the instance to create the code to generate an html page will also work to create the same document in Latex, or the opposite. LATEXReport can be used to create presentations with beamer.

There are other useful classes designed to read the JPL (\cite*{Pickett1998}) and CDMS (\cite*{Muller2001}) molecular databases, execute system commands, work with RSS feeds, use the printer, or interact with the system clipboard. System commands are used in LATEXReport to optionally compile a tex file and show the resulting file in the default PDF viewer of the system. This kind of commands are system dependent, but JPARSEC hides this complexity providing a cross-system implementation.

\subsection{Charting and rendering}

The package jparsec.graph contains classes to create different kind of charts. For this purpose some external libraries are integrated in JPARSEC, with classes aimed to simplify and unify the interaction with them. JFreeChart\footnote{http://www.jfree.org/jfreechart/} is used for classical charts (dispersion, bar, and pie charts), SGT\footnote{https://www.pmel.noaa.gov/epic/java/sgt/} is used for 2d surface charts, and 3d charts are created with VISAD\footnote{http://www.ssec.wisc.edu/~billh/visad.html}, SurfacePlotter\footnote{https://github.com/ericaro/surfaceplotter}, JMathPlot\footnote{https://github.com/yannrichet/jmathplot}, and Jzy3d\footnote{http://www.jzy3d.org/}. There is also a class to facilitate the use of ditaa\footnote{http://ditaa.sourceforge.net/}, a package to create flow diagrams. Some of the libraries were modified to add features such as subscripts and superscripts, useful to show units.

Charts of the sky can be created with the class jparsec.graph.SkyChart, that can be consider a basic planetarium. The sky rendering itself is implemented in the class RenderSky inside the chartRendering sub-package. There are other classes here to render eclipse maps, visibility maps of artificial satellites, or realistic renderings of the planets. There are many example programs in JPARSEC that show how to use these features and the large amount of options available to create a specific chart.

The package jparsec.io.image contains the class Picture to manipulate images (supporting among other common operations scaling, rotation, and changes in contrast, brightness, and color levels), and the class Draw to create simple drawings. Charts created with JFreeChart, SGT, ditaa, Draw, and also sky and planetary renderings use the native Java class Graphics2D. This class can be replaced with the Graphics2D implementation of the external libraries FreeHEP or iText (also provided with JPARSEC), allowing to export these charts to the vector graphics formats pdf, eps, and svg.

\begin{figure}[!h]
    \centering
    \includegraphics[width=0.65\textwidth]{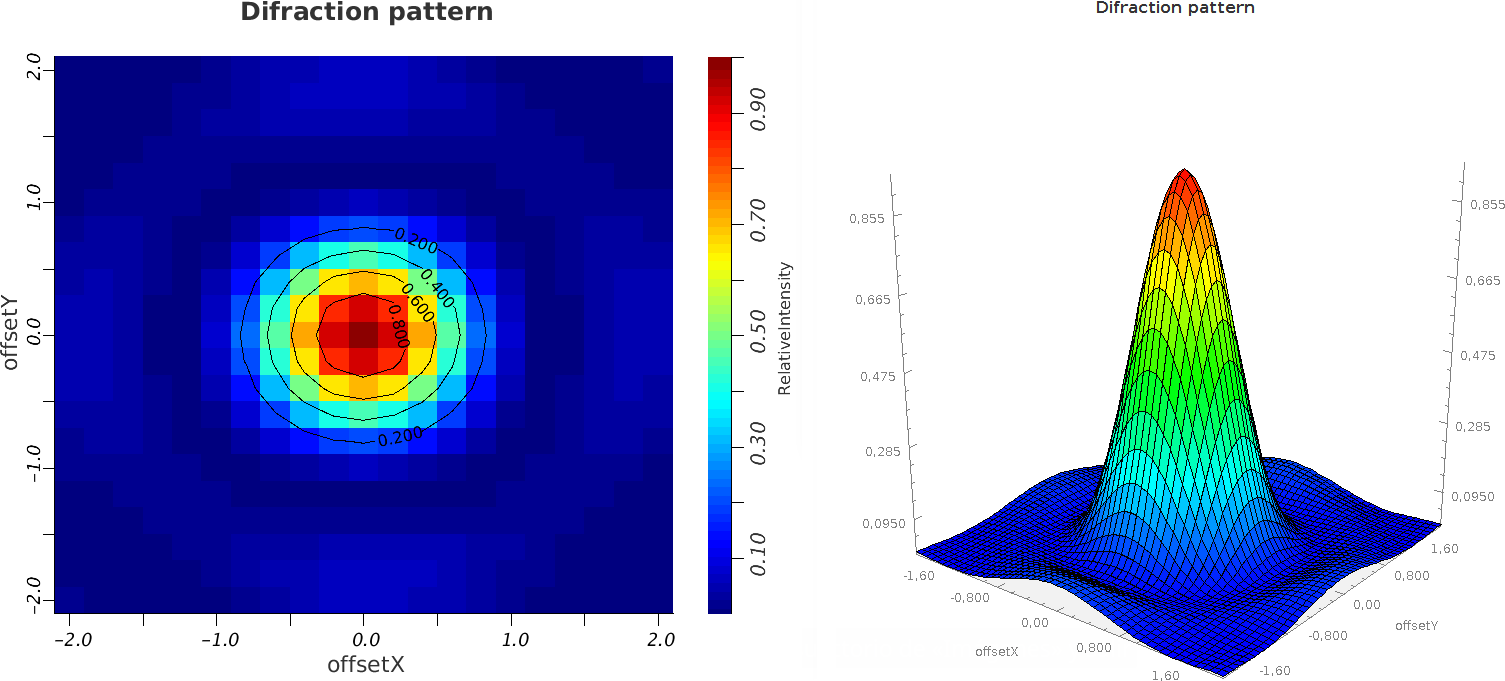}
    \caption{Charts resulting from the code listing \ref{list2}. They are interactive: scale, rotation angle, and other properties can be changed within them.}
    \label{fig:charts}
\end{figure}

\begin{lstlisting}[caption={Code example to generate charts with the SGT and SurfacePlotter libraries showing the difraction pattern of a telescope. See Fig. \ref{fig:charts}.},captionpos=b,label={list2}]
// Obtain the difraction pattern of a Newton 20cm telescope
TelescopeElement telescope = TelescopeElement.NEWTON_20cm;
int field = 2; // arcsec
double dat[][] = Difraction.pattern(telescope, field);
GridChartElement chart = new GridChartElement("Difraction pattern", "offsetX", "offsetY", "RelativeIntensity", GridChartElement.COLOR_MODEL.BLUE_TO_RED, new double[] {-field, field, -field, field}, dat, new double[] {0, 0.2, 0.4, 0.6, 0.8, 1.0}, 800);
 
// Plot using SGT
CreateGridChart cs0 = new CreateGridChart(chart);
cs0.showChartInSGTpanel(true);
 
// Plot using SurfacePlotter
CreateSurface3D cs1 = new CreateSurface3D(chart);
cs1.show(800, 800, chart.title, true);
\end{lstlisting}

The code listing \ref{list2} and Fig. \ref{fig:charts} show how to plot the difraction pattern of a telescope with the SGT and SurfacePlotter libraries. Other figures below show additional examples of charts created with JFreeChart, and sky and planetary rendering.

\subsection{Other utilities}

The package jparsec.model contains some classes to calculate dust extinction and opacity, compute gas temperatures and column densities using the rotational diagram method, or use a Java implementation of RADEX (\cite*{vandertak2007}). The models in this package are used in different tools and programs, but are a subset of those previously available and later moved to external specific tools.

The class jparsec.io.Zip can be used to compress and uncompress zip files or directories. This feature is used in the jparsec.util.Update class to update the orbital elements of minor bodies and other data present in the dependencies of JPARSEC. These dependencies are files with .jar extension that are compressed in zip format, containing in some cases external libraries and in other cases text files that require updates periodically. Having all these data compressed saves a lot of disk space.

Another useful class is jparsec.util.DataBase. This class can be used to store whatever data is needed in a program, specially useful when the program is complex and spread in multiple files. The stored data (arrays for instance) are identified with a name and optionally also by the thread that stores the data. Later, the data can be recovered or deleted from any part of the program. Data can be stored with a given lifetime, after that it will be deleted.

The package jparsec.io.device and its sub-package provides an implementation of the communication protocols of several amateur telescopes, a wrapper to use gPhoto\footnote{http://gphoto.org/} to control digital cameras, and some algorithms useful for basic reduction of optical astronomical images. Telescope control and astrophotography are already supported, but some work is pending.

The class jparsec.util.Logger is used to send messages to the user with different significance levels, that can be shown in the console or written to a text file with optional filters to those levels. Exceptions and warnings during the execution of a program are handled by the class jparsec.util.JPARSECException. Programs are by default stopped after an error, but this can be overridden to try to continue the execution. Warnings are by default shown in the console, but they can be occulted or the program stopped when a warning is thrown.

The class jparsec.util.Module is used to keep the current version of any of the dependencies of the JPARSEC library or its associated tools. One of the tools developed is a manager that automatically updates the library or the tools when a new version is uploaded.

Other features provided with JPARSEC includes the possibility of uploading and downloading files using the SFTP protocol and even to send or read emails. They are provided by additional external libraries, but as with charts JPARSEC includes classes that are wrappers to use them with easy. The intention of these wrappers is to reduce the amount of code lines needed for most operations with the library, inviting to code against JPARSEC instead of all these libraries. An additional advantage is the possibility of changing the underlying dependencies by other better libraries in the future, without requiring changes in any of the programs using these features.

\section{Projects developed}

\begin{figure*}
    \centering
    \includegraphics[width=1.0\textwidth]{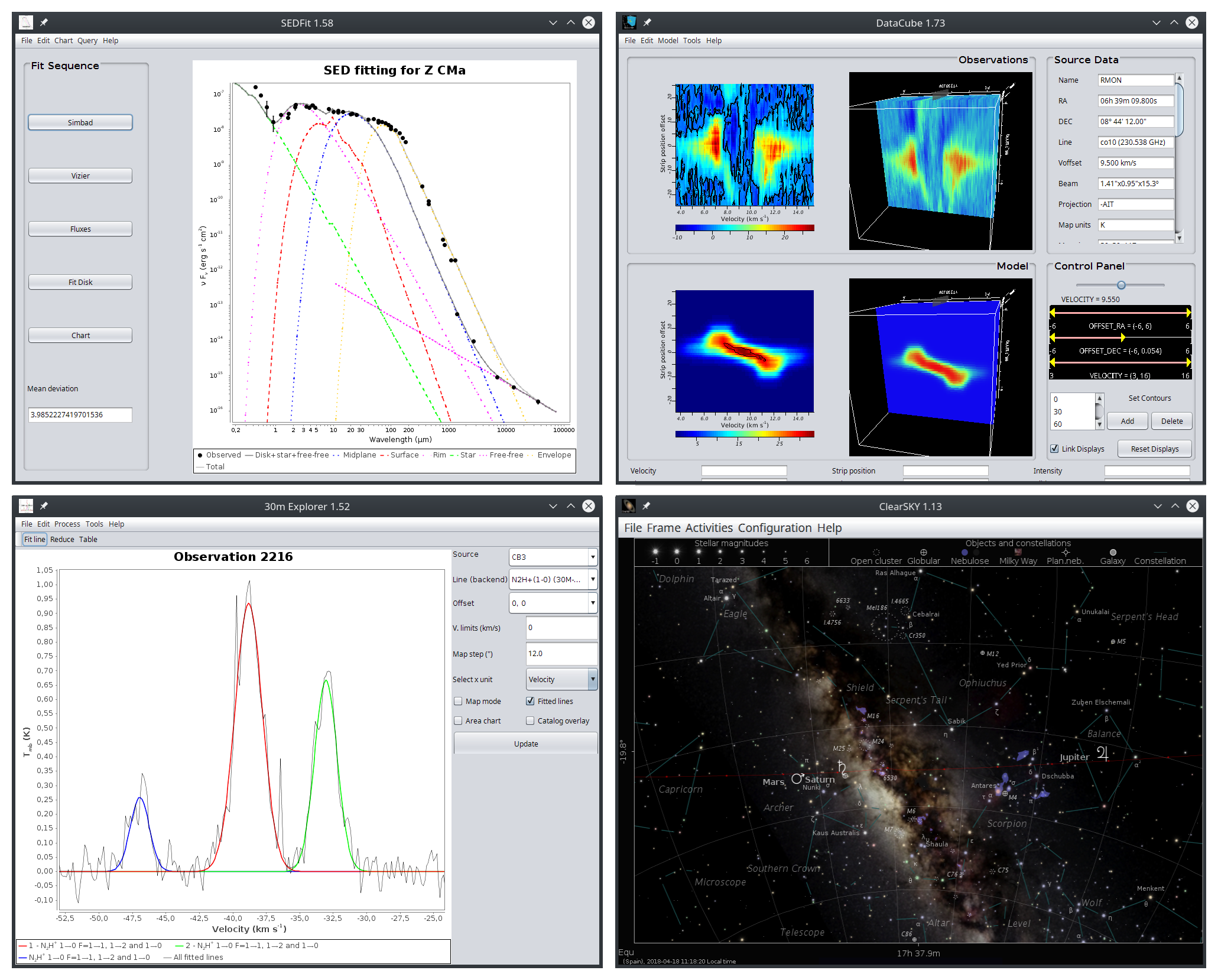}
    \caption{Screenshots of different tools with graphical user interfaces. Up-Left: SEDFit showing the SED fitting for Z CMa from UV to cm wavelengths. Up-Right: DataCube, showing a position-velocity diagram along the major axis of the R Mon circumstellar disk for the CO 2-1 transition (observations above the model, and the interactive 3d cube cuts on the right). Down-Left: 30mExplorer showing fits to the N$_2$H$^+$ 1-0 transitions for CB3. Down-Right: ClearSky planetarium component, in which the Android planetarium is based.}
    \label{fig:Models}
\end{figure*}

In this section I present the main projects developed based on JPARSEC. The tools with graphical user interfaces (GUIs) are mainly models that were developed in parallel with JPARSEC (see Fig. \ref{fig:Models}). They are like satellites of the JPARSEC library, using a minimum amount of code lines and a complete GUI for a more comfortably operation. Later, with a more stable and performance wise core, even more laborious projects (Fig. \ref{fig:Projects}) were implemented. The amount of Java code lines in these projects is around 190 000, also released under GPL license except the ephemerides server and the GUI of the Android planetarium.

\subsection{Modeling tools}

There are nine tools with a GUI based on JPARSEC, most of them freely available for downloading. I present here only the three most relevant ones used for modeling across different Papers. In some cases the modeling work was carried out through the GUI itself, and in more complex cases in batch mode directly from Java code.

\subsubsection{SEDFit: SED fitting model for circumstellar disks}

SEDFit is a tool designed to fit the spectral energy distribution (SED) of protostars from UV to cm wavelengths. For this purpose different models were implemented and integrated together in the tool: the Kurucz model (\cite*{Buser1992}) for stellar atmospheres, the circumstellar disk model by \cite{Dullemond01}, a simple black body based model for the protostellar envelope (if any), and an additional component for the free-free emission in cm wavelengths. This was the first model that successfully fitted the entire SED of several sources (\cite*{AlonsoAlbi2008,Alo09,Boissier2011}), considering the different beams or resolutions of the continuum observations carried out with a variety of instruments: VLA and PdBI interferometers, 30m and JCMT single dishes, and different continuum surveys in UV, optical, and IR wavelengths.

The circumstellar disk model by \cite{Dullemond01} is the main implementation in this tool. It was developed in Java from the original IDL code, performing the corresponding changes for our use case, and the required tests. 

The GUI show some buttons on the left. The process of fitting consists on selecting each of them to first solve the source with Simbad, then to obtain the photometry (automatically for a number of surveys available at Vizier), then to add custom photometry or fluxes at other wavelengths, and then to fit the circumstellar disk and envelope parameters. Among the options to fit the disk it is possible to remove the inner rim present in the model to use a pure \cite{Chiang2001} disk model, or to change the grain size distribution and composition.

SEDFit can run scripts to repeat the calculations for a given range in each parameter, selecting the best fitting model. It is also possible to export charts as GILDAS scripts, showing how the SED varies with changes in different parameters, for instance for the grain size distribution.

\subsubsection{DataCube: a radiative transfer model}

DataCube is a radiative transfer model that can compute, for different available geometries, the output emission for a given transition using the local thermodynamic equilibrium (LTE) approximation, and the large velocity gradient (LVG) method as implemented in RADEX. The input parameters are the size and scaling properties of the geometry (sphere, cylinder, toroid, or double cone, among others), the distance, inclination, and position angle of the source, the line transition (selected from the JPL and CDMS databases), the gas properties (density, temperature, abundance, velocity, and velocity dispersion profiles), and a few additional parameters that affect the behavior of the model. The cylinder geometry can be used to model a circumstellar disk, using the scaling factors to reduce the altitude of the cylinder, and, if necessary, different values of this altitude for the inner and outer radius (flared disk). The profiles for the gas properties are set as string functions dependent on a number of available variables (or other custom ones that can be created) that are later dynamically evaluated in each point during the computation of the model. With this approach the model is not dependent on the particular profile desired for a given model, and different profiles (with static gas or in Keplerian rotation) can be selected by changing the input formula.

Depending on the figure and its orientation the radiative transfer will be computed in one or several cuts of the figure along the line of sight, evaluating the input profiles to compute the gas conditions in each region. The radiative transfer is then computed in each cell (of a given configurable depth) from behind the figure toward the observer, integrating the emission and considering the optical depth. This is extended to the entire grid to obtain a spatial map of the model, resulting in a 3d cube with the velocity in the third axis. The observations (a GILDAS lmv cube) can optionally be loaded to directly compare cuts and strips, and in this case the cube resolution and beam are automatically adjusted. The tool shows two charts for the observations and the model: a 2d or surface chart with the map for a given velocity channel, the integrated intensity, or a position-velocity cut, and a 3d interactive cube that can be rotated and cut in right ascension, declination, and velocity (see top-right panel of Fig. \ref{fig:Models}). When any of these charts is modified, the corresponding one above or below it is automatically changed accordingly in case a link checkbox is selected. These kind of charts can also be created and integrated in other programs using some classes included in the jparsec.graph package. The spectra for the observations and model at each grid position (selected by moving the mouse on the surface chart) can be compared on the screen.

As with the SEDFit tool there are a number of additional features: scripts can also be used for fitting, cubes can be rescaled or clipped, results can be exported to Latex, and cubes, strips, and individual spectra can be exported to GILDAS format. pyGildas interaction is available in case the goodness of the fitting should be evaluated with GILDAS. 

DataCube has been used in different Papers: \cite{Fuente2006,Alo10}; and Alonso-Albi et al. (2018) (accepted in A\&A). In \cite{Fuente2006} and Alonso-Albi et al. (2018) the input density profile was a rather long formula selected to preserve the mass of the circumstellar disk from any possible change in the size of the cylinder geometry. A similar approach was applied to the velocity profile, as function in this case of the mass of the star R Mon itself. Hence the use of these symbolic profiles naturally produced as a result the appearance of secondary input parameters in the model.

\subsubsection{30mExplorer: line fitting tool for radioastronomy}

This tool reads GILDAS spectra (old format only) and assists in the process of reducing them, fitting Gaussians, and identifying lines. The results can be exported to Latex format, including custom tables with selected columns (with the errors of the Gaussian parameters properly formatted). There are some additional tools, among them to stitch spectra, reduce on the fly maps, create 30m spectra from custom data, create setup charts for 30m observations, or transform spectra in the new 30m format to the old format. 30mExplorer can also export charts to scripts in GILDAS format, with different levels of detail depending on the intensity peak of the lines. 

Thanks to the GUI, exploring spectra and fitting Gaussians in 30mExplorer is easy and fast. The fitting routine uses the Java scientific library by Michael Thomas Flanagan. When developing this tool several algorithms were tested against each other, and this library proved to be the best option. The fitting values were tested against GILDAS, showing very similar values for the error parameters of the fittings when the signal to noise ratio of the data is very good, and slightly larger values of these errors when the line intensities are of a few $\sigma$.

\begin{figure*}
    \centering
    \includegraphics[width=0.60\textwidth]{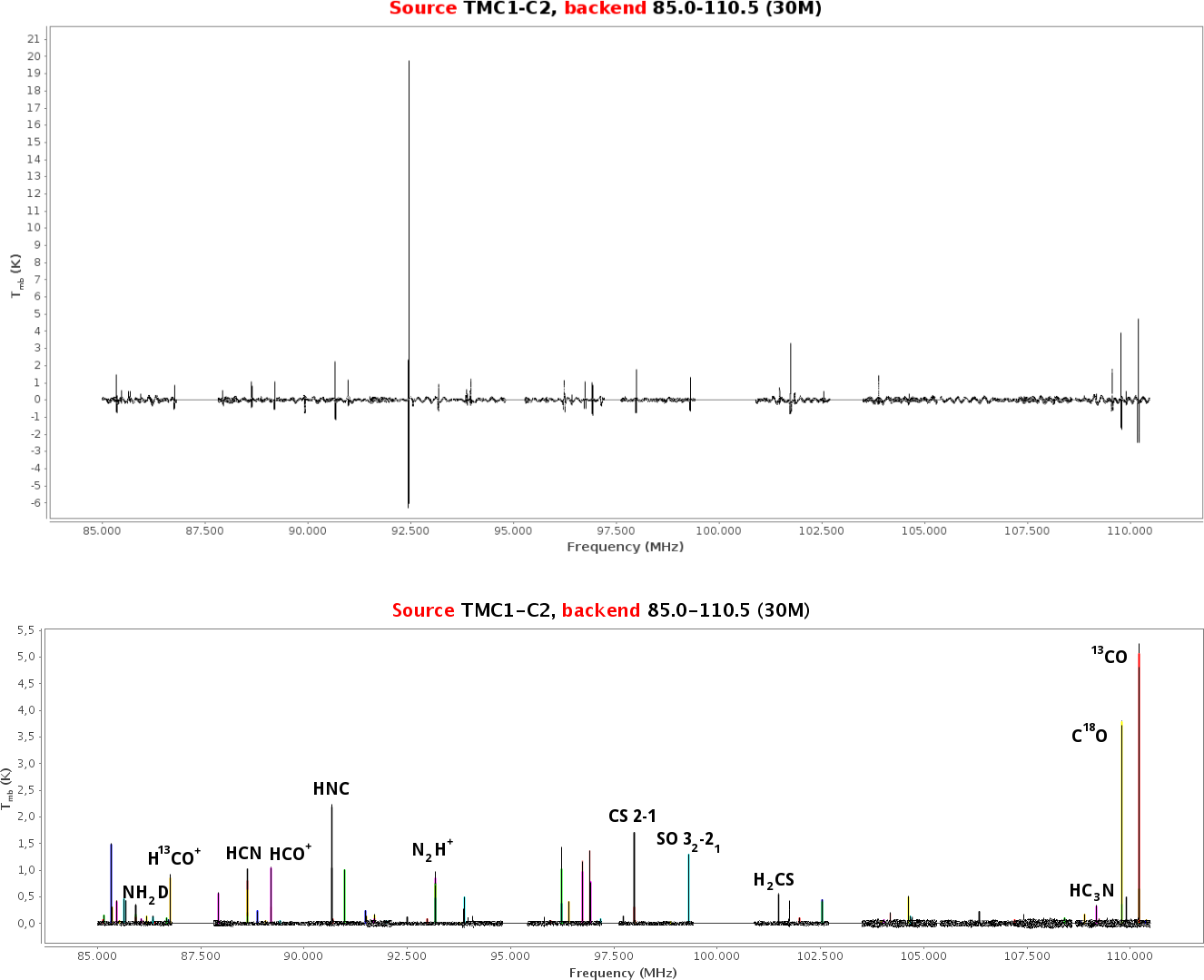}
    \includegraphics[width=0.39\textwidth]{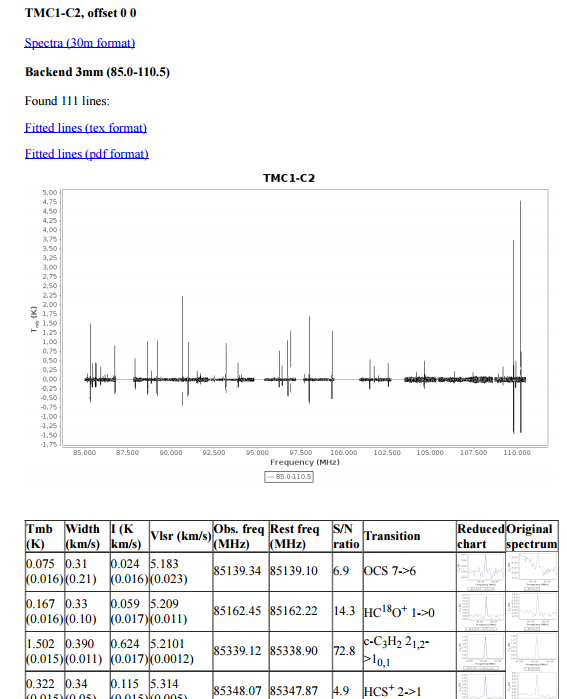}
    \caption{Left: Stitch of the spectra between 85 and 110 GHz for the source TMC1-C2 as observed in the GEMS large project. The upper panel shows the raw spectrum while the bottom panel shows the reduced one (the lines with negative intensities due to the frequency switching are substracted) and some lines identified. Right: Screenshot of a web page showing the results of the reduction, in this case without subtracting the lines with negative intensities.}
    \label{fig:GEMS}
\end{figure*}

The ongoing 30m large program GEMS\footnote{Gas phase Elemental abundances in Molecular cloudS (GEMS) is being carried out with the IRAM-30m single dish radiotelescope.} (PI Asunci\'on Fuente) is using the underlying algorithms present in JPARSEC for reducing 
the data. In GEMS we use a completely automatic pipeline that uses the pyGildas extension as a final step to stitch the spectra with GILDAS, but the rest of the process (including data reduction, line fitting and identification, text reports and data upload) is performed with JPARSEC (see Fig.~\ref{fig:GEMS}). The implemented pipeline reduces the observations in frequency switching mode in different steps. A custom catalog of molecular lines is used to mask the channels in which a line could be present, then a dynamic baseline subtraction is applied to flatten the spectrum. The set of observations are then stacked, and the line fitting and identification algorithm is applied. The line fitting process searches for lines in the catalog above the 3 $\sigma$ level, extracting the different velocity components that could be present by means of an iterative method based on subtracting and fitting again each component several times. Only components above the 3 $\sigma$ level are considered. In addition, unidentified lines are also fitted if any appear above a 5 $\sigma$ level. These lines are latter identified (if possible) and added to the catalog, and then the entire process is eventually repeated to improve the reduction of these new lines in all positions. Since there are many target positions in this project it is easy to recognize unidentified lines, since they appear repeatedly. The error rate of the automatic pipeline when searching for lines or fitting them is only around 5\%. These cases of wrong lines or fittings are easy to identify since the resulting line fittings are exported to tables with charts showing the spectrum before the reduction and after the reduction and line fitting. 

30mExplorer is not flexible enough to properly work with observations in frequency switching mode, and, although it can reduce some data automatically, it is not designed to be a fully automatic pipeline. 30mExplorer was used by \cite{Ginard2012}.

\subsection{Android planetarium}

An Android planetarium based on JPARSEC is available for downloading\footnote{https://play.google.com/store/apps/details?id=jparsec.androidlite}. It was first released three years ago, but based on six years and several thousands of hours of work. To implement this planetarium the first step was to improve as much as possible the performance and to reduce the memory requirements for sky rendering in JPARSEC, using the VisualVM profiler tool. This process started when JPARSEC was entering the stable phase, and extended to the entire library during a few months to greatly improve (by a factor 10) the performance in many operations with the library. The performance requirements to reach a reasonable speed in Android devices led to a replace of some intrinsic drawing operations by other custom ones.

The sky rendering pipeline in JPARSEC was modified to use the interface Graphics located in the jparsec.graph.chartRendering package. This interface performs generic drawing calls that later are executed with a particular implementation of the interface. There is an implementation of the interface for desktop programs (AWTGraphics), and a different one for Android devices (AndroidGraphics). At pixel level there are some differences in the way certain functions, like filling a rectangle with color and many others, behave in Java for desktop and in Android. The interface allows to correct for these differences and to map all drawing operations to get the same exact result at pixel level in those two platforms. This approach can be used to generate any other kind of drawings using the same code for both platforms.

\begin{figure*}
    \centering
    \includegraphics[width=1.0\textwidth]{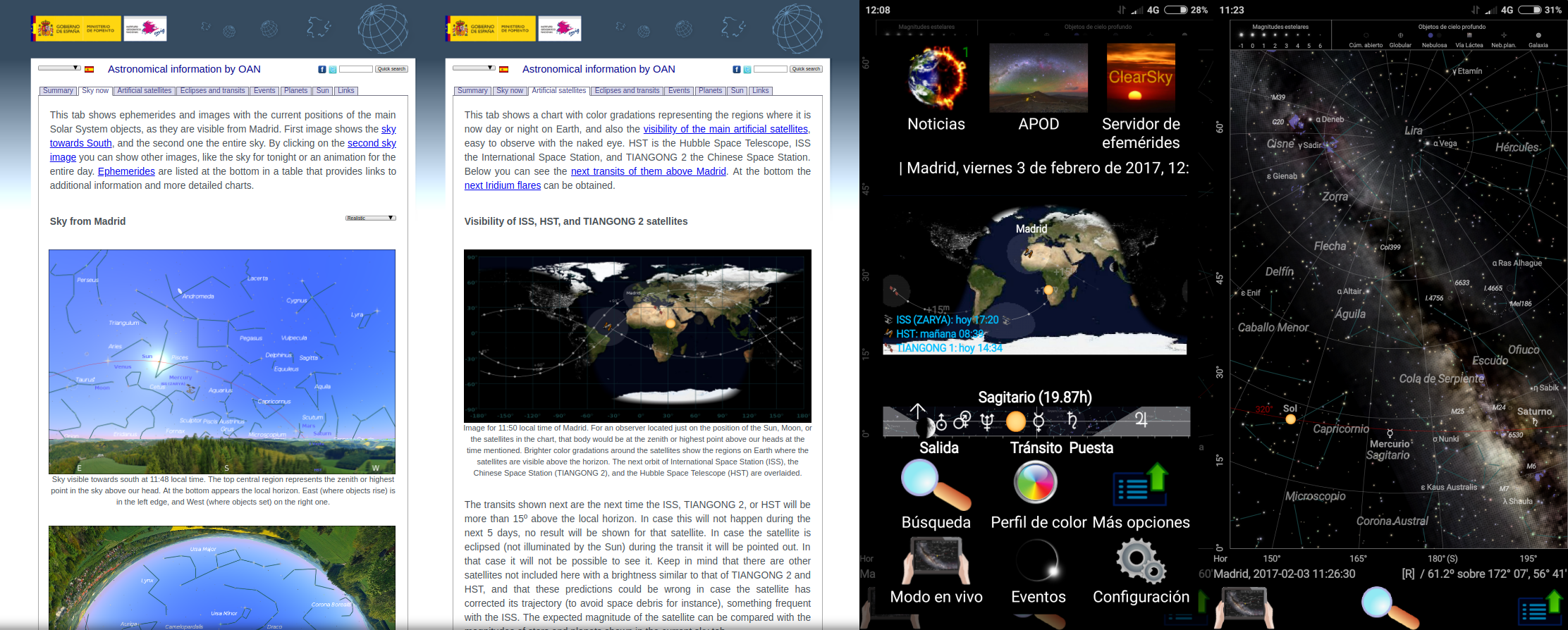}
    \caption{Left: Screenshots of the sky and artificial satellites tabs of the ephemerides server. Right: Screenshots (in Spanish) of the main menu and sky window of the Android planetarium.}
    \label{fig:Projects}
\end{figure*}

The implementation of the planetarium itself was rather difficult and laborious due to the differences between desktop and Android platforms, and the particular fixes required to make it work over the huge variety of devices available. The help of the developer community at stackoverflow\footnote{https://stackoverflow.com/} was essential for the success in this project. The program (see Fig. \ref{fig:Projects}, right panel) is rated 4.4/5\footnote{It is possible to pay for false downloads and 5-star ratings. I have not followed this practice.}, making it one of the best rated planetariums in the Google Play Store.

\subsection{Ephemerides server}

A web ephemerides server\footnote{http://www.oan.es/servidorEfem/eindex.php for the English version.} is another laborious application of JPARSEC available. This project started before the Android planetarium, but was finished in its current design later. There are other comparable web resources available, among them the previously mentioned JPL Horizons, the IMCCE ephemerides server, or the calcsky\footnote{https://www.calsky.com} portal. The goal of this ephemerides server based on JPARSEC is to provide a more understandable, wider, and visually rich information, useful both for the general public and amateur astronomers. 

The main page first tries to use geolocalization to obtain the user position, without the need of any input data. When this fails there is an optional list to select a location. An additional form\footnote{http://www.oan.es/servidorEfem/form.php} is also offered to request specific ephemerides for a given object, date, and location of the observer.

The information in the ephemerides server is loaded at once, and organized in tabs. In each of them the relevant information is showed with some explanations, and all data (tables and charts) are generated on the fly and downloaded in few seconds. The information showed covers most of the astronomical calculations available in the JPARSEC library. The sky and planetary charts stands out in quality above other resources, and the ephemerides for planets, comets, and asteroids are always updated, with orbital elements downloaded automatically. Since this ephemerides server was developed mainly for the general public the planetary ephemerides are based on the analytical theory by Moshier previously mentioned, with no option to use any of the JPL integrations.

The JPARSEC library and the rest of projects, including the Android planetarium and the ephemerides server, supports different languages. Currently Spanish and English translations for the strings present in the projects are officially supported, and a first version of an Italian translation is offered in JPARSEC.

The current implementation of the ephemerides server uses PHP to call the underlying Java programs. This approach is more simple compared to Java servlets, but since this is a slow language there is a great performance penalty. The main page loads in about ten seconds, that could be reduced to around two given the performance of JPARSEC. A migration to Java servlets and a redesign of the page (for mobile devices) are pending tasks.

\subsection{Other applications and outreach}

In addition to the main web page of the JPARSEC library, described in the next section, other web pages show additional applications of JPARSEC developed in the past ten years. There is a complete star atlas showing how to render and export sky charts to PDF format and include them in a Latex document (beamer presentation) with some text and tables. This project was developed as a proof of concept and recently greatly improved by Angelo Nicolini in his atlas\footnote{The Bungee Star Atlas is a complete atlas based on JPARSEC. The project is available at https://bitbucket.org/anjiloh/bungeeskyatlas/. Angelo is also the author of the Italian translation of JPARSEC.}. The blog page mentions other applications of JPARSEC over the years: charts for the annual of the Observatorio Astron\'omico de Madrid, 3d videos rendered pixel by pixel and frame by frame, or different collaborations with amateur and professional astronomers, engineers, and software developers.

\begin{figure*}
    \centering
    \includegraphics[width=0.495\textwidth]{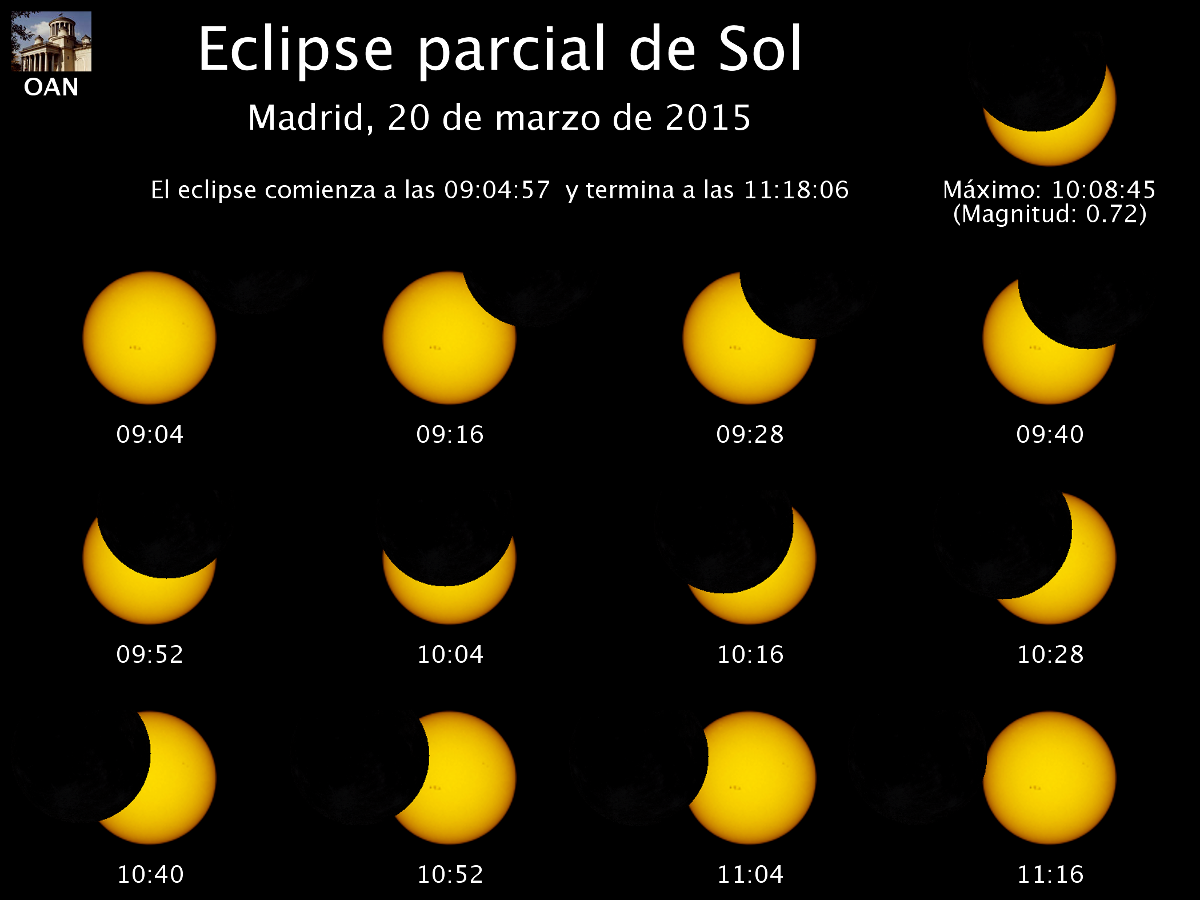}
    \includegraphics[width=0.495\textwidth]{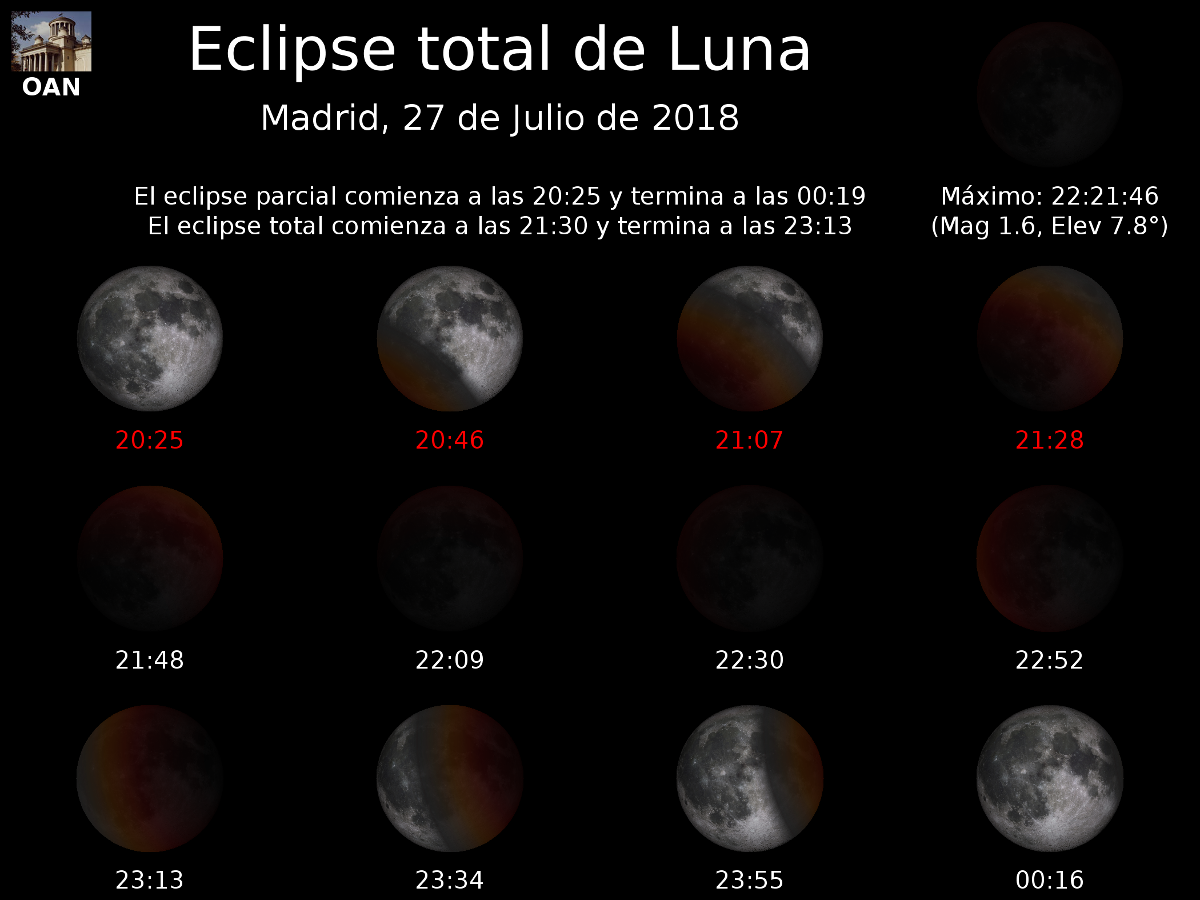}
    \caption{Rendering examples used for outreach. Left: Sequence of the partial solar eclipse of 2015 from Madrid, Spain, showing text in Spanish language. Right: Total lunar eclipse of 2018 also from Madrid. First half of the sequence will not be visible, and this is indicated with red labels.}
    \label{fig:outreach}
\end{figure*}

JPARSEC has been used as a core resource to generate charts for outreach. In particular, charts showing the sequence of solar and lunar eclipses (see Fig. \ref{fig:outreach}) have appeared in different media, from web pages\footnote{http://www.oan.es/eclipse2015/} and social networks to newspapers and television. Both the basic atlas and the eclipse sequence are among the many useful code examples of JPARSEC.

\section{Documentation and how to start using JPARSEC}

In this section I describe the different web resources available to install JPARSEC and start using the source code and modeling tools. The source code is distributed under the GPL license.

\subsection{Downloading and installing}


The main page of the JPARSEC library\footnote{http://conga.oan.es/\%7Ealonso/doku.php?id=jparsec} provides the basic information on how to download and start using it, but there are several ways to do that. One option, first presented, is to use an installer to download JPARSEC, its dependencies, and later any of the models available. The installer will run on Linux, Mac, and Windows operating systems with the only condition of having Java 1.6 installed. For Linux and Windows the installer asks to download a particular Java Runtime Environment (JRE) 1.6 with additional libraries required for 3d visualization, a feature only used in the DataCube tool. Although not required, it can be very useful to select both JREs in case of later copying the entire installation directory to a different operating system (or even executing the software from an external portable disk memory), since everything will work without any change. The installer will copy basic scripts to launch a manager tool from the console, and a clickable file for Windows systems. Once the manager is executed the required libraries will be downloaded, and in a second launch the manager will list the available tools, allowing to install and execute any of them. The launching java command admits additional parameters to directly execute a specific program during startup and even to load an input file on it.

A second option is to install JPARSEC to an IDE, to work with the source code. The main JPARSEC page describes how to do that for the Eclipse IDE\footnote{http://www.eclipse.org/downloads/}, although the basic process is similar for any other. There is still a third method: to download the latest source code from the JPARSEC repository at bitbucket\footnote{https://bitbucket.org/talonsoalbi/jparsec/}. The repository was created in collaboration with Carlo Dapor, who also contributed with many Java specific code changes and tests to improve the quality of the JPARSEC library. This option is more advanced, but will provide the latest source code. In both cases many files need to be downloaded to solve dependencies. Since this is done automatically when using the installer it is recommended to use it before trying to work with the source code, although the dependencies will also be downloaded from the repository with the Maven\footnote{https://maven.apache.org/} tool.

The main page of JPARSEC provides other links: a special version of the library for Android developers, or a complete tutorial with instructions for the Eclipse IDE, which is described below.

\subsection{Library tests and example programs}

The JPARSEC internal tests are available from the source code repository at bitbucket. The test directory contains 160 files with 15000 lines of code that check most of the basic functionalities of JPARSEC, distributed in packages with the same structure of the library. Some of these code pieces are written as JUnit\footnote{https://junit.org/junit5/} tests.

This test directory contains an additional package jparsec.test, composed of several Java classes that executes an additional 302 tests for JPARSEC. There are tests related to coordinate conversions, calendars, different time scales, error propagation, analytical theories for ephemerides computations, or astronomical events. Some of them are fine-grained tests recovered independently from external sources: ephemerides results from the Astronomical Almanac, historical occultations of stars by planets, and mutual events between the satellites of Jupiter and Saturn, observed or predicted by other trusted sources. The tests are listed and formatted in a text file, including the exact date and time for the test, the objects involved, and when relevant the expected exact mathematical result of the ephemerides for a particular analytical theory and reduction algorithms. The class jparsec.test.Test reads this file and execute the computations with JPARSEC to check if the test passes or fails. The execution of this class is a ritual before the release of a new version of JPARSEC at the main web page, ensuring a consistent quality of the library.

There are also 65 additional and more elaborated example programs (around 21 000 lines of code) that cover many possible applications of JPARSEC. These examples are available for downloading from the main web page of the JPARSEC project. In my own research I developed around two hundred of additional code pieces not available for downloading. All this work together makes JPARSEC a thoroughly tested library.

\subsection{Tutorial to use the source code of JPARSEC}

A complete tutorial is published online\footnote{http://conga.oan.es/\%7Ealonso/doku.php?id=jparsec\%5Fwiki}, covering from the installation instructions for Eclipse to a detailed description, with code examples, of most of the features of JPARSEC. The contents are also available in PDF format, with different versions optimized for desktop and tablets.

The goal of this page is to introduce JPARSEC and to provide help to get used to the library at source code level. For this purpose the tutorial starts explaining how to install the Eclipse IDE itself and how to configure a project for JPARSEC and run the examples previously mentioned on it. The tutorial goes through progressive levels of complexity explaining with examples many of the features of JPARSEC (ephemerides, charting, data manipulation and fitting, Virtual Observatory, among other sections), while giving advice to some good general development practices. This tutorial has been followed with success, at least in the first steps for running the examples and creating some charts, even by people with no previous knowledge of Java. 

\section{Summary and conclusions}

In this Paper I present the JPARSEC library. It is not a new library, but a Java based project started twelve years ago. The library is developed with strict code conventions, extensive documentation inside and outside the code, and a large collection of code examples and tests. It has great potential for astronomers and since it is distributed under the GPL license others can contribute to extend or improve it. Some modeling tools and associated projects are also described, completing a volume of work above half a million lines of code.

The development of JPARSEC started as a tool to improve the productivity and safety of the scientific work, but also to eventually use it as a supporting tool for different levels of outreach. These general goals required a library integrating many different kind of calculations, with complex charts and renderings, and a flexible and productive language suitable for a wide range of applications (scientific programs, web projects, or even portable devices that were inexistent when the project started). The main conclusion after twelve years is that to base the research on this library has proven to be extremely productive, and its features has been used for many different purposes and exciting collaborations that were impossible to foresee when the development started.

JPARSEC is now in a stable phase. There is still a large list of features that remain to be implemented or completed, but none of them is specially important. The same applies to the modeling tools and other projects presented in this Paper.

\bibliography{jparsec}

\end{document}